\documentclass[11pt,prd,superscriptaddress,nofootinbib]{revtex4-2}
\usepackage{graphicx}
\usepackage{float}
\usepackage{placeins}
\usepackage[dvipsnames]{xcolor}
\usepackage{booktabs}
\usepackage[colorlinks=true, linkcolor=blue, citecolor=blue, urlcolor=blue]{hyperref}
\begin{document}

\title{Bound State Solutions of the Relativistic Finite-difference Equation for the Ring-shaped Quesne Oscillator Potential}


\author{Sh.M.Nagiyev}
\email{shakir.m.nagiyev@gmail.com}
\affiliation{Institute of Physics, Ministry of Science and Education,\\
H. Javid Avenue 131, AZ1073, Baku, Azerbaijan}

\author{Narmin Nasibova}
\email{n.nesibli88@gmail.com}
\affiliation{Institute of Physics, Ministry of Science and Education,\\
H. Javid Avenue 131, AZ1073, Baku, Azerbaijan}
\affiliation{Center for Theoretical Physics, Khazar University,\\
41 Mehseti Str., AZ1096, Baku, Azerbaijan}

\author{V. A. Tarverdiyeva}
\email{vefa.tarverdiyeva@mail.ru}
\affiliation{Institute of Physics, Ministry of Science and Education,\\
H. Javid Avenue 131, AZ1073, Baku, Azerbaijan}
\affiliation{Sumgait State University,\\
43rd District, Baku Street 1, AZ5008, Sumgait, Azerbaijan}

\author{G. H. Guliyeva}
\email{gulnara.quliyeva@sdu.edu.az}
\affiliation{Sumgait State University,\\
43rd District, Baku Street 1, AZ5008, Sumgait, Azerbaijan}
\begin{abstract}
\end{abstract}

\maketitle
\thispagestyle{empty}

\vspace{2cm}
\centerline{\bf Abstract} 
\vskip 4mm
		{\Large \bf }
		\vskip 0.5 cm

\author{
Institute of Physics, Ministry of Science and Education,\\
H. Javid Avenue, 131, AZ 1143, Baku, Azerbaijan\\
\texttt{shakir.m.nagiyev@gmail.com}}
\maketitle

We solve exactly the relativistic finite-difference equation for the quantum three-dimensional ring-shaped Quesne oscillator potential. Our investigation is based on a finite-difference version of relativistic quantum mechanics. So-called relativistic configurational $\bf r$-space is a key concept here. We show that the radial wavefunctions and angular wavefunctions are expressed through the continuous dual Hahn polynomials and Jacobi polynomials, respectively. A discrete energy spectrum has been found. The radial wave functions and energy spectrum have the correct nonrelativistic limit. We also build a dynamical symmetry group $SU(1,1)$ for the radial part of the equation of motion, which allows us to find the energy spectrum purely algebraically.


\clearpage
\tableofcontents

\vspace{1cm}

\section{Introduction}
\label{int}
In quantum mechanics (relativistic and nonrelativistic) often arise the problems related to finding the energy levels and the wave functions of a particle moving in a certain potential force field \cite{ref1,ref2,ref3}. A fundamental equation of motion of the nonrelativistic quantum mechanics is the Schr\"{o}dinger equation. It describes physical phenomena in molecular physics, atomic and nuclear physics, quantum chemistry, etc., occurring at low energies \cite{ref1,ref2}. The description of physical phenomena occurring at high energies should be based on relativistic wave equations \cite{ref2,ref3}. If the interaction potential is not enough to create particle--antiparticle pairs, we can apply the Klein-Gordon equation to the treatment of zero-spin particle and apply the Dirac equation to that of a $\frac{1}{2}$ spin particle. Note that the Klein-Gordon equation and the Dirac equation are the most frequently used equations for the description of the particle dynamics in relativistic quantum mechanics. We also note that, in addition to this, there is another, namely, the finite-difference version of relativistic quantum mechanics, developed in the works (see \cite{ref4,ref5,ref6,ref7,ref8,ref9,ref10,ref11,ref12,ref13} and references there).

In many works using various central and non-central (or ring-shaped) potentials the problems of both non-relativistic (e.g., \cite{ref14,ref15,ref16,ref17,ref18,ref19,ref20,ref21,ref22,ref23,ref24,ref25}) and relativistic (e.g., \cite{ref26,ref27,ref28,ref29,ref30,ref31}) bound states and scattering states are studied. It should be especially noted that non-central or ring-shaped potentials are widely used in various fields of theoretical physics, for example, in nuclear and atomic physics, in the physics of molecules, and also in quantum chemistry. Such potentials are used in quantum chemistry to describe the properties of the organic ring-shaped molecules such as benzene, and in nuclear physics to study the interactions of a deformed pair of nuclei. One needs to note that Quesne in published work \cite{ref15} introduces a new ring-shaped oscillator potential. The Quesne potential is defined by

\begin{equation}
V_Q(r,\vartheta) = \frac{\sigma^2}{2\eta^2}\left(\frac{r^2}{b_0^2} + q\eta^4 \frac{b_0^2}{r^2 \sin^2\vartheta}\right)\varepsilon_0,
\label{1}
\end{equation}
where $b_0 = \left(\hbar / (m_0\omega\right)^{1/2}$, $\varepsilon_0 = \frac{1}{2}\hbar\omega$, and $\eta, \sigma, q$ are three dimensionless positive parameters. For $q=0$ and $\sigma^2 = 2\eta^2$, we obtain the harmonic oscillator potential. The exact solution of the relativistic finite-difference wave equation with the ring-shaped Coulomb potential was found in \cite{ref30}.

The purpose of the present paper is to generalize the exactly solvable nonrelativistic Quesne ring-shaped oscillator potential (\ref{1}) to the relativistic finite-difference quantum mechanics case and construct a dynamical symmetry group for the radial equation.

The remainder of this paper is structured as follows. \autoref{int} provides the introduction. \autoref{2s} presents the description of finite-difference relativistic quantum mechanics. \autoref{3s} introduces the finite-difference relativistic ring-shaped oscillator model. In \autoref{4s} and \autoref{5s}, the solutions to the angular and radial equations are derived, respectively. \autoref{6s} examines the non-relativistic limit, while \autoref{7s} establishes the dynamical symmetry group associated with the radial wave equation. Finally, \autoref{v} presents the numerical results, and \autoref{con} offers the concluding remarks.

\section{The finite-difference relativistic quantum mechanics}
\label{2s}
The key concept in the finite-difference version of relativistic quantum mechanics is the notion of the three-dimensional relativistic configurational $\bf r$-space. The corresponding canonically conjugate momentum $\bf p$-space is the three-dimensional Lobachevsky space realized on the upper sheet of the mass hyperboloid
\begin{equation}
p_0^2 - \mathbf{p}^2 = m_0^2 c^2, \quad p_0 > 0.
\end{equation}
Both $\bf r$- and $\bf p$-spaces are related to each other by the relativistic Fourier transform
\begin{equation}
\psi(\mathbf{r}) = \frac{1}{(2\pi\hbar)^{3/2}} \int \xi(\mathbf{p},\mathbf{r})\, \psi(\mathbf{p})\, d\Omega_p,
\end{equation}
where $d\Omega_p = m_0 c\, dp / p_0$ is the relativistic three-dimensional volume element in the momentum Lobachevsky space and the function $\xi(\mathbf{p},\mathbf{r})$ is the relativistic plane waves
\begin{equation}
\xi(\mathbf{p},\mathbf{r}) = \left(\frac{p_0 - \mathbf{p}\mathbf{n}}{mc}\right)^{-1 - ir/\lambdabar}.
\label{wa}
\end{equation}
Here, $\mathbf{r} = r\mathbf{n}$, $0 \leq r < \infty$, $\mathbf{n} = (\sin\vartheta\cos\varphi, \sin\vartheta\sin\varphi, \cos\vartheta)$ is the unit vector along the radius vector, and $\lambdabar = \hbar/(m_0 c)$ is the Compton wavelength of the particle. The relativistic plane waves ( \ref{wa}) obey the finite-difference free Schr\"{o}dinger equation
\begin{equation}
(H_0 - E_p)\xi(\mathbf{p},\mathbf{r}) = 0, \quad E_p = cp_0,
\label{5}
\end{equation}
where the finite-difference operator
\begin{equation}
H_0 = m_0 c^2 \left[\cosh(i\lambdabar\partial_r) + i\frac{\lambdabar}{r}\sinh(i\lambdabar\partial_r) + \frac{\bf L^2}{2(m_0 cr)^2}e^{i\lambdabar\partial_r}\right]
\end{equation}
is the free Hamiltonian and ${\bf L^{2}} = -\hbar^2 \Delta_{\vartheta,\varphi}$ is the square of the angular momentum operator
\begin{equation}
{\bf L^2} = -\hbar^2\left[\frac{1}{\sin\vartheta}\partial_\vartheta(\sin\vartheta\,\partial_\vartheta) + \frac{1}{\sin^2\vartheta}\partial_\varphi^2\right].
\end{equation}
The equation for the wave function of the relative motion of two scalar particles with the interaction potential $V({\bf r})$ in the relativistic configurational $\bf r$-space can be obtained from (\ref{5}) and has the form
\begin{equation}
[H_0 + V({\bf r})]\psi(\mathbf{r}) = E\psi(\mathbf{r}).
\label{2.7}
\end{equation}
In the nonrelativistic limit we have
\begin{equation}
\lim_{c\to\infty} \xi(\mathbf{p},\mathbf{r}) = e^{i\mathbf{p}\mathbf{r}/\hbar}.
\end{equation}

\section{The finite-difference relativistic ring-shaped Quesne oscillator model}
\label{3s}
We now consider a finite-difference relativistic ring-shaped oscillator model with the potential
\begin{equation}
V(r) = \left[\frac{1}{2}m_0\omega^2(r + i\lambdabar)^2 + \frac{\alpha}{r^2\sin^2\vartheta}\right]e^{i\lambdabar\partial_r},
\label{eq:3.1}
\end{equation}
where $\alpha = (q\sigma^2\eta^2\hbar^2)/(4m_0)$ is a positive parameter.
This potential is a generalization of the Quesne potential (\ref{1}) to the case of finite-difference relativistic quantum mechanics.

The potential ( \ref{eq:3.1}) has a correct nonrelativistic limit (\ref{1}):
\begin{equation}
\lim_{c\to\infty} V({\bf r}) = \frac{1}{2}m_0\omega^2 r^2 + \frac{\alpha}{r^2\sin^2\vartheta}.
\end{equation}
The operator (\ref{eq:3.1}) is Hermitian with respect to a scalar product
\begin{equation}
(\psi_1, \psi_2) = \int \psi_1^*({\bf r})\,\psi_2({\bf r})\,dr,
\end{equation}
i.e.\ $(V\psi_1, \psi_2) = (\psi_1, V\psi_2)$. Here, the functions $\psi_1({\bf r})$ and $\psi_2({\bf r})$ vanish at the points $r=0$ and $r=\infty$ together with all their derivatives.

We want to find a solution to equation (\ref{2.7}) with potential( \ref{eq:3.1}). To do this, we define $\psi(\mathbf{r}) = \frac{1}{r}R({\bf r})$, then we obtain the following equation for the wave function $R{\bf (r)}$:
\begin{equation}
\Bigl[m_0 c^2 \cosh(i\lambdabar\partial_r) + \frac{{\bf L^2}}{2m_0 r^{(2)}}e^{i\lambdabar\partial_r} \nonumber\\
+ \left(\frac{1}{2}m_0\omega^2 r^{(2)} + \frac{\alpha}{r^{(2)}\sin^2\vartheta}\right)e^{i\lambdabar\partial_r} - E\Bigr]R({\bf r}) = 0,
\label{eq:3.5}
\end{equation}
where $r^{(\delta)} = (i\lambdabar)^{\delta}\Gamma(\delta -\frac{ir}{\lambdabar})/\Gamma( -\frac{ir}{\lambdabar})$ is a generalized degree. This equation allows the separation of variables. Since the Hamiltonian of Eq. (\ref{eq:3.5}) commutes with the operator $\hat{L}_z = -i\hbar\partial_\varphi$, we can look for the wave function in the form
\begin{equation}
R(\mathbf{r}) = R(r)\,F(\vartheta)\,\frac{1}{\sqrt{2\pi}}e^{im\varphi},
\label{eq:3.6}
\end{equation}
where $m = 0, \pm1, \pm2, \ldots$ is the usual magnetic quantum number. Substitution of (\ref{eq:3.6}) into Eq.\ (\ref{eq:3.5}) leads to a set of finite-difference and second-order differential equations
\begin{equation}
\left[\cosh(i\partial_\rho) + V_{eff}(\rho) - \varepsilon\right]R(\rho) = 0, ~ V_{eff}(\rho)=\left(\frac{1}{2}\omega_0^2\rho^{(2)} + \frac{\Lambda}{\rho^{(2)}}\right)e^{i\partial_\rho},
\label{eq:3.7}
\end{equation}

\begin{equation}
\left[\partial_\vartheta^2 + \cot\vartheta\,\partial_\vartheta - \frac{M^2}{\sin^2\vartheta} + \Lambda\right]F(\vartheta) = 0,
\label{eq:3.8}
\end{equation}
where $\rho = r/\lambdabar$ is a dimensionless variable, $M^2 = m^2 + (2m_0\alpha)/\hbar^2$, $\Lambda$ is a separation constant, and $\rho^{(2)} = \rho(\rho+i)$.

The value for the parameter $\Lambda$ will be determined from the angular Eq. (\ref{eq:3.8}). In addition, we have introduced the following notations $\omega_0 = \hbar\omega/(m_0 c^2)$, $\varepsilon = E/(m_0 c^2)$. Obviously, when $\alpha = 0$, we have $\Lambda = l(l+1)$.

\section{The solutions of the angular equation}
\label{4s}
We investigate the solutions of the angle-dependent equation (\ref{eq:3.8}) \cite{ref30}. In terms of the new variable $x = \cos\vartheta$, we obtain the following equation for the function $F(\vartheta) \equiv F(x)$:
\begin{equation}
\left[\partial_x^2 - \frac{2x}{1-x^2}\partial_x + \frac{c_0 - c_2 x^2}{(1-x^2)^2}\right]F(x) = 0,
\label{eq:4.1}
\end{equation}
where
\begin{equation}
c_0 = \Lambda - M^2, \quad c_2 = \Lambda.
\label{eq:4.2}
\end{equation}
We look for the solution of eq.\ (\ref{eq:4.1}) in the form \cite{ref29}
\begin{equation}
F(x) = \varphi(x)\,y(x), \quad \varphi(x) = (1-x)^A(1+x)^B,
\label{eq:4.3}
\end{equation}
where $\varphi(x)$ is the weight part, and $y(x)$ is the polynomial part of the angular wave function $F(x)$. From the condition of finiteness of $F(x)$ at points $x = \pm 1$ it follows that the parameters $A$ and $B$ must satisfy the inequalities $A \geq 0$, $B \geq 0$. Then for the function $y(x)$ we obtain the following equation
\begin{equation}
y''(x) - \frac{2\delta + 2(s+1)x}{1-x^2}y'(x) + \frac{\gamma_2 x^2 + \gamma_1 x + \gamma_0}{(1-x^2)^2}y(x) = 0.
\label{eq:4.4}
\end{equation}
Here we use the notations $\delta = A - B$, $s = A + B$, and also
\begin{equation}
\gamma_2 = -c_2 + s + s^2, \quad \gamma_1 = 2\delta s, \quad \gamma_0 = c_0 - s + \delta^2.
\label{eq:4.5}
\end{equation}

Now let us choose the parameters $A$ and $B$ so that the relation $\gamma_2 x^2 + \gamma_1 x + \gamma_0 = \lambda(1-x^2)$ is satisfied, i.e.\ $\gamma_2 = -\lambda$, $\gamma_1 = 0$, $\gamma_0 = \lambda$, where $\lambda = \mathrm{const}$. From these equalities we find the parameters $A$, $B$ and $\lambda$:
\begin{equation}
A = B = \frac{1}{2}\sqrt{c_2 - c_0} = \frac{1}{2}|M|,
\label{eq:4.6}
\end{equation}
\begin{equation}
\lambda = c_2 - |M| - M^2.
\label{eq:4.7}
\end{equation}

Now, Eq. (\ref{eq:4.4}) takes the form
\begin{equation}
(1-x^2)y''(x) - (2|M|+2)x\,y'(x) + \lambda\,y(x) = 0.
\label{eq:4.9}
\end{equation}
Comparing it with the equation for Gegenbauer polynomials $\bar{y}(x) = C_k^{(\beta)}(x)$, $k = 0,1,2,\ldots$ \cite{ref32}:
\begin{equation}
(1-x^2)\bar{y}''(x) - (2\beta+1)x\,\bar{y}'(x) + k(k+2\beta)\bar{y}(x) = 0,
\label{eq:4.10}
\end{equation}
we get $\beta = |M| + \frac{1}{2}$ and
\begin{equation}
\lambda \equiv \lambda_k = k(k + 2|M| + 1).
\label{eq:4.11}
\end{equation}
Gegenbauer polynomials satisfy the following orthogonality condition:
\begin{equation}
\int_{-1}^{1}(1-x^2)^{\beta - 1/2} C_k^{(\beta)}(x)\,C_{k'}^{(\beta)}(x)\,dx = \frac{\pi\,\Gamma(k+2\beta)\,2^{1-2\beta}}{\{\Gamma(\beta)\}^2(k+\beta)\,k!}\,\delta_{kk'},
\end{equation}
where $\beta > -\frac{1}{2}$ and $\beta \neq 0$.

Thus, we conclude that the solution of equation (\ref{eq:4.9}) $y(x) \equiv y_k(\cos\vartheta)$ is expressed through Gegenbauer polynomials, i.e.
\begin{equation}
y_k(\cos\vartheta) = C_k^{(|M|+1/2)}(\cos\vartheta).
\label{eq:4.12}
\end{equation}
We can now write an expression for the total angular wave function in terms of the angle $\vartheta$:
\begin{equation}
F_{km}(\cos\vartheta) = c_{km}\,(\sin\vartheta)^{|M|}\\C_k^{(|M|+1/2)}(\cos\vartheta).
\label{eq:4.13}
\end{equation}
The normalization constant $c_{km}$ is found from the orthonormality condition
\begin{equation}
\int_0^\pi F_{km}(\cos\vartheta)\,F_{k'm}(\cos\vartheta)\,\sin\vartheta\,d\vartheta = \delta_{kk'}.
\label{eq:4.14}
\end{equation}
It equals to
\begin{equation}
c_{km} = 2^{-|M|}\sqrt{\frac{(k+|M|+\frac{1}{2})\,k!\,\Gamma(k+2|M|+1)}{\Gamma(k+|M|+1)}}.
\label{eq:4.15}
\end{equation}
Let us now emphasize that from the two formulas (\ref{eq:4.9}) and (\ref{eq:4.11}) for the value $\lambda$ we find the following quantized expression for the separation parameter $\Lambda \equiv \Lambda_k = L(L+1)$, where $L = k + |M|$. It is easy to see that when $\alpha = 0$ we have $l = k + m$.

Let us emphasize that due to the relationship between the Gegenbauer polynomials and the associated Legendre function of the first kind \cite{ref33}:
\begin{equation}
P_n^m(\cos\vartheta) = (-2)^m\left(\frac{1}{2}\right)_m (\sin\vartheta)^{|M|}\;C_{n-m}^{(m+1/2)}(\cos\vartheta),
\label{eq:4.16}
\end{equation}
the angular wave function (\ref{eq:4.13}) coincides with the corresponding result of \cite{ref15}. Therefore, as in \cite{ref15}, in the case we are considering, there are also two integrals of motion:
\begin{equation}
W_1 ={\bf L^2} + \frac{2m_0\alpha}{\sin^2\vartheta} = -\frac{\hbar^2}{\sin\vartheta}\partial_\vartheta(\sin\vartheta\,\partial_\vartheta) - \frac{1}{\sin^2\vartheta}(\hbar^2\partial_\varphi^2 - 2m_0\alpha),
\label{eq:4.17}
\end{equation}
\begin{equation}
W_2 = L_z^2 + 2m_0\alpha = -\hbar^2\partial_\varphi^2 + 2m_0\alpha.
\label{eq:4.18}
\end{equation}

\section{The solutions of the radial equation}
\label{5s}
We will obtain the solutions of Eq.\ (\ref{eq:3.7}). To solve Eq.\ (\ref{eq:3.7}), we choose $R(\rho)$ as \cite{ref11}
\begin{equation}
R(\rho) = (-\rho)^{(\mu_k)}\,M_{\nu_k}(\rho)\,\Omega(\rho).
\label{eq:5.1}
\end{equation}
Here the separated factors
\begin{equation}
\rho^{(\mu_k)} = i^{\mu_k}\frac{\Gamma(\mu_k - i\rho)}{\Gamma(-i\rho)}, \quad M_{\nu_k}(\rho) = \omega_0^{i\rho}\,\Gamma(\nu_k + i\rho)
\label{eq:5.2}
\end{equation}
determine the asymptotic behaviour of $R(\rho)$ at the points $\rho = 0$ and $\rho = \infty$ respectively, where
\begin{equation}
\mu_k = \frac{1}{2} + \frac{1}{2}\sqrt{1 + \frac{2}{\omega_0^2}\!\left(1 - \sqrt{1 - 4\omega_0^2\Lambda_k}\right)},
\end{equation}
\begin{equation}
\nu_k = \frac{1}{2} + \frac{1}{2}\sqrt{1 + \frac{2}{\omega_0^2}\!\left(1 + \sqrt{1 - 4\omega_0^2\Lambda_k}\right)}
\label{eq:5.3}
\end{equation}
are the real parameters. It follows that
\begin{equation}
\Lambda_k \leq \left(\frac{m_0 c^2}{2\hbar\omega}\right)^2.
\label{eq:5.4}
\end{equation}
This condition imposes an upper limit on the values of the ``orbital'' quantum number $k$.
The function $\Omega(\rho)$ then satisfies the following finite-difference equation:
\begin{equation}
\left[(\mu_k + i\rho)(\nu_k + i\rho)e^{-i\partial_\rho} - (\mu_k - i\rho)(\nu_k - i\rho)e^{i\partial_\rho}\right]\Omega(\rho) = \frac{2\varepsilon}{\omega_0}\,i\rho\,\Omega(\rho).
\label{eq:5.5}
\end{equation}

Let us compare this equation with the equation for continuous dual Hahn polynomials $y(x) = S_n(x^2; a, b, c)$ at $c = \frac{1}{2}$ \cite{ref32}:
\begin{equation}
\left[(a+ix)(b+ix)e^{-i\partial_x} - (a-ix)(b-ix)e^{i\partial_x}\right]y(x) = (2a + 2b + 4n)\,ix\,y(x).
\label{eq:5.6}
\end{equation}
From the comparison we get that $a = \mu_k$, $b = \nu_k$, $\varepsilon = \omega_0(2n + \mu_k + \nu_k)$, where $n = 0,1,2,\ldots$ is the radial quantum number. As a result we have
\begin{equation}
E_{nk} = \hbar\omega(2n + \mu_k + \nu_k).
\label{eq:5.7}
\end{equation}
This gives the quantization rule for the energy levels of the three-dimensional finite-difference relativistic ring-shaped oscillator model under consideration. Thus, the polynomial solution of the finite difference equation (\ref{eq:5.5}) is expressed through the continuous dual Hahn polynomials:
\begin{equation}
\Omega(\rho) \equiv \Omega_{nk}(\rho) = S_n\!\left(\rho^2;\,\mu_k,\nu_k,\frac{1}{2}\right).
\label{eq:5.8}
\end{equation}
Note that the continuous dual Hahn polynomials are defined with relation \cite{ref32}:
\begin{equation}
S_n(x^2;a,b,c) = (a+b)_n(a+c)_n\;
{}_3F_2\!\left(
\begin{array}{c}
-n,\ a+ix,\ a-ix \\
a+b,\ a+c
\end{array}
; 1
\right).
\label{eq:5.9}
\end{equation}

If $a$, $b$ and $c$ are positive except possibly for a pair of complex conjugates with positive real parts, then the continuous dual Hahn polynomials satisfy the following orthogonality condition \cite{ref32}:
\begin{eqnarray}
\frac{1}{2\pi}\int_0^\infty \left|\frac{\Gamma(a+ix)\,\Gamma(b+ix)\,\Gamma(c+ix)}{\Gamma(2ix)}\right|^2 S_n(x^2;a,b,c)\,S_m(x^2;a,b,c)\,dx \nonumber\\
= \Gamma(n+a+b)\,\Gamma(n+a+c)\,\Gamma(n+b+c)\,n!\;\delta_{nm}.
\label{eq:5.10}
    \end{eqnarray}

We can now write down an expression for the radial wave function $R(\rho)$ (\ref{eq:5.1}), corresponding to the energy levels (\ref{eq:5.7}):
\begin{equation}
 R(\rho)\equiv R_{nk}(\rho) = \tilde{c}_{nk}\,(-\rho)^{(\mu_k)}\,M_{\nu_k}(\rho)\,S_n\!\left(\rho^2;\,\mu_k,\nu_k,\frac{1}{2}\right).
\label{eq:5.11}
\end{equation}
The orthogonality condition for the function (\ref{eq:5.11}) easily follows from the condition (\ref{eq:5.10}):
\begin{equation}
\int_0^\infty R_{nk}^*(\rho)\,R_{mk}(\rho)\,d\rho = \delta_{nm}.
\label{eq:5.12}
\end{equation}
From here we obtain the following expression for the normalization constant:
\begin{equation}
\tilde{c}_{nk} = 2^{1/2}\left[n!\,\Gamma(n+\mu_k+\nu_k)\,\Gamma\!\left(n+\mu_k+\frac{1}{2}\right)\Gamma\!\left(n+\nu_k+\frac{1}{2}\right)\right]^{-1/2}.
\label{eq:5.13}
\end{equation}
Hence, the total wave function takes the form:
\begin{equation}
\psi_{nkm}(\mathbf{r}) = \tilde{c}_{nk}\,c_{km}\,\frac{1}{r}\,\left(-\frac{r}{\lambdabar}\right)^{(\mu_k)}\!M_{\nu_k}(\rho)\,S_n\!\left(\left(\frac{r}{\lambdabar}\right)^2;\mu_k,\nu_k,\frac{1}{2}\right)(\sin\!\vartheta)^{|M|}\; C_k^{(|M|+1/2)}(\cos\vartheta)\,\frac{1}{\sqrt{2\pi}}e^{im\varphi}.
\label{eq:5.14}
\end{equation}
Wave function (\ref{eq:5.14}) satisfies the normalization condition $\int \psi_{nkm}^*({\bf r})\,\psi_{ n'k'm'}({\bf r})d {\bf r} = \delta_{nn'}\delta_{kk'}\delta_{mm'}.
$

\section{The non-relativistic limit case}
\label{6s}
It can be easily shown that formula (\ref{eq:5.7}) for the energy spectrum and formula (\ref{eq:5.11}) for the radial wave function of the relativistic ring-shaped Quesne oscillator have a correct nonrelativistic limit, i.e.\ they in the nonrelativistic limit $c \to \infty$ pass into the corresponding formulas of the nonrelativistic ring-shaped oscillator. For this, it is sufficient to take into account the following limit and asymptotic relations \cite{ref11}:
\begin{equation}
\lim_{c\to\infty}\lambdabar^{\mu_k} = \frac{1}{2} + \frac{1}{2}\sqrt{1+4\Lambda_k} = L+1, \quad \lim_{c\to\infty}\!\left(\nu_k - \frac{1}{\omega_0}\right) = \frac{1}{2},
\label{eq:6.1}
\end{equation}
\begin{equation}
\lim_{c\to\infty}\lambdabar^{\mu_k}(-\rho)^{(\mu_k)} = (-r)^{L+1}, \quad \lim_{c\to\infty} M_{\nu_k}(\rho) \cong \sqrt{2\pi}\exp\!\left(\frac{1}{\omega_0}\ln\frac{1}{\omega_0} - \frac{1}{\omega_0} - \frac{1}{2}\lambda_0^2 r^2\right),
\label{eq:6.2}
\end{equation}
\begin{equation}
\lim_{c\to\infty}\frac{\omega_0^n}{n!}S_n\!\left(\rho^2;\mu_k,\nu_k,\frac{1}{2}\right) = L_n^{L+1/2}(\lambda_0^2 r^2).
\label{eq:6.3}
\end{equation}
Here $\lambda_0 = \sqrt{m_0\omega/\hbar}$ and $L_n^a(z)$ are the generalized Laguerre polynomials \cite{ref32,ref33}.

\section{Dynamical symmetry group for the radial wave equation}
\label{7s}
In this section we will construct a dynamical symmetry group for the radial equation (\ref{eq:3.7}). We will show that this group is the group $SU(1,1)$. Its generators $K_0$, $K_1$ and $K_2$ satisfy the commutation relations \cite{ref34}:
\begin{equation}
[K_0, K_1] = iK_2, \quad [K_2, K_0] = iK_1, \quad [K_1, K_2] = -iK_0.
\label{eq:su11_comm}
\end{equation}
In terms of the lowering and raising operators $K^\pm = K_1 \pm iK_2$ they are written as
\begin{equation}
[K_0, K^\pm] = \pm K^\pm, \quad [K^-, K^+] = 2K_0.
\label{eq:raising_lower}
\end{equation}
The Casimir operator is
\begin{equation}
C_2 = K_0^2 - K_1^2 - K_2^2 = K_0^2 - \frac{1}{2}(K^-K^+ + K^+K^-).
\end{equation}
We denote the states of a positive discrete series $D^+(s)$ as $|n,s\rangle$ such that $K_0|n,s\rangle = (n+s)|n,s\rangle$ and $C_2|n,s\rangle = s(s-1)|n,s\rangle$, $s$ being the Bargmann index, $s > 0$ and $n = 0,1,2,3,\ldots$ Action of operators $K^\pm$ on states $|n,s\rangle$ is given by formulas:
\begin{equation}
K^-|n,s\rangle = \sqrt{n(n+2s-1)}\,|n-1,s\rangle, \quad K^+|n,s\rangle = \sqrt{(n+1)(n+2s)}\,|n+1,s\rangle.
\label{eq:6.4}
\end{equation}

To construct the dynamical symmetry group for the radial equation of motion (\ref{eq:3.7}), we note that the operators \cite{ref35}:

\begin{eqnarray}
    a= \frac{1}{\sqrt{2}}\left[e^{-\frac{i}{2}\partial_\rho} - \omega_0\,e^{\frac{i}{2}\partial_\rho}(\mu_k + i\rho)\!\left(1 + \frac{\nu_k}{i\rho}\right)\right], \label{eq:6.5a}\\
a^+ = \frac{1}{\sqrt{2}}\left[e^{-\frac{i}{2}\partial_\rho} - \omega_0(\mu_k - i\rho)\!\left(1 - \frac{\nu_k}{i\rho}\right)e^{\frac{i}{2}\partial_\rho}\right] \label{eq:6.5b}
\end{eqnarray}
factorize the Hamiltonian of equation (\ref{eq:3.7}), i.e.
\begin{equation}
H = m_0 c^2\left[a^+ a + \omega_0(\mu_k + \nu)\right].
\label{eq:6.6}
\end{equation}
However, the operators $a^+$, $a$ and $H$ do not form a closed algebra, since
\begin{equation}
[a, a^+] = \frac{1}{2}\omega_0\!\left(1 + \frac{\mu_k\nu_k}{\rho^2 + \frac{1}{4}} + \omega_0 Q\,e^{i\partial_\rho}\right),
\end{equation}
\begin{equation}
Q = \mu_k + \nu_k - \frac{1}{4} + \mu_k\nu_k\!\left[-\frac{(\mu_k-1)(\nu_k-1)}{\rho^{(2)}} + \frac{\mu_k\nu_k}{(\rho + \frac{i}{2})^{(2)}}\right].
\label{eq:6.7}
\end{equation}
We now look for the decreasing and increasing operators in the form:
\begin{eqnarray}
b = i\rho\!\left[\sqrt{2}\,e^{-\frac{i}{2}\partial_\rho}a - \frac{H}{m_0 c^2} + \omega_0(\mu_k+\nu_k)\right] + \frac{H}{2m_0 c^2} + \omega_0\mu_k\nu_k, \label{eq:6.8a}\\
b^+ = -i\!\left[\sqrt{2}\,a^+e^{\frac{i}{2}\partial_\rho} - \frac{H}{m_0 c^2} + \omega_0(\mu_k+\nu_k)\right]\rho + \frac{H}{2m_0 c^2} + \omega_0\mu_k\nu_k. \label{eq:6.8b}
\end{eqnarray}
It can now be verified by direct verification that
\begin{equation}
[H, b] = -2\hbar\omega\,b, \quad [H, b^+] = 2\hbar\omega\,b^+.
\label{eq:6.9}
\end{equation}
Their commutator is equal to
\begin{equation}
[b, b^+] = \frac{\omega_0}{m_0 c^2}\,H\!\left\{1 + \frac{2}{\omega_0^2}\!\left[\left(\frac{H}{m_0 c^2}\right)^2 - 1\right]\right\}.
\label{eq:6.10}
\end{equation}
If we define the generalized momentum operator  $P$ according to the formula
\begin{equation}
P = -\frac{i}{c}[\rho, H],
\label{eq:6.11}
\end{equation}
then the operators $b$, $b^+$ can be written in compact form:
\begin{equation}
b = \frac{1}{2\omega_0}\left[\left(\omega_0\rho + \frac{i}{m_0 c}P\right)^2 - \frac{2\Lambda}{\rho^2+1}\right],
\end{equation}
\begin{equation}
b^+ = \frac{1}{2\omega_0}\left[\left(\omega_0\rho - \frac{i}{m_0 c}P\right)^2 - \frac{2\Lambda}{\rho^2+1}\right].
\label{eq:6.12}
\end{equation}
From definition (\ref{eq:6.11}) we find an expression for the generalized momentum operator:
\begin{equation}
P = -m_0 c\!\left[\sinh(i\partial_\rho) + \frac{1}{2}\omega_0^2\rho^{(2)}e^{i\partial_\rho} + \frac{\Lambda}{\rho^{(2)}}e^{i\partial_\rho}\right].
\label{eq:6.13}
\end{equation}

From (\ref{eq:6.10}) it follows that to construct the algebra of dynamical symmetry of the system under consideration we introduce the following operators:
\begin{equation}
K^- = b\,f^{-1/2}(H), \quad K^+ = f^{-1/2}(H)\,b^+,
\label{eq:6.14}
\end{equation}
where
\begin{equation}
f(H) = \left[\frac{H}{m_0 c^2} + \omega_0(\mu_k - \nu_k - 1)\right]\!\left[\frac{H}{m_0 c^2} - \omega_0(\mu_k - \nu_k + 1)\right].
\label{eq:6.15}
\end{equation}

It is now easy to check that the operators $K^-$, $K^+$ and $K_0 = H/2\hbar\omega$ satisfy commutation relations in space $\{R_{nk}(\rho)\}_{n=0}^\infty$, i.e.
\begin{equation}
[K_0, K^\pm] = \pm K^\pm, \quad [K^-, K^+]R_{nk} = 2K_0 R_{nk},
\label{eq:6.16}
\end{equation}
 In this case, the Casimir operator is equal to:
\begin{equation}
K^2 = \frac{\mu_k+\nu_k}{2}\!\left(\frac{\mu_k+\nu_k}{2}-1\right), \quad s = \frac{\mu_k+\nu_k}{2}.
\label{eq:6.17}
\end{equation}
Therefore, the eigenvalues of the Hamiltonian $H = 2\hbar\omega K_0$ are bounded below and equal $E_{nk} = \hbar\omega(2n + \mu_k + \nu_k)$, $n = 0,1,2,3,\ldots$, and eigenfunctions $\{R_{nk}\}$ forms a basis for an irreducible representation $D^+\!\left(\frac{\mu_k+\nu_k}{2}\right)$ of the group $SU(1,1)$.

In conclusion, we note that the action of the operators $b$ and $b^+$ on their own functions of $H$ are determined by formulas:
\begin{equation}
b\,R_{nk} = b_{nk} R_{n-1,k}, \quad b^+R_{nk} = b_{n+1,k}R_{n+1,k},
\end{equation}
\begin{equation}
b_{nk} = 2\omega_0\sqrt{n(n+\mu_k+\nu_k)\!\left(n+\mu_k-\frac{1}{2}\right)\!\left(n+\nu_k-\frac{1}{2}\right)}.
\label{eq:last_6.15}
\end{equation}
From group considerations it is clear that arbitrary states $R_{nk}$ of the system under consideration can be obtained by $n$-multiple actions on the ground state $R_{0k}$, i.e.
\begin{equation}
R_{nk} = N_{nk}(b^+)^n R_{0k} = c_{nk}(-\rho)^{(\mu_{k})}\omega_0^{i\rho}\,\Gamma(\nu_{k}+i\rho)\,S_n\!\left(\rho^2;\mu_k,\nu_k,\frac{1}{2}\right),
\label{eq:6.16b}
\end{equation}
where
\begin{equation}
N_{nk}^{-1} = b_{1k} b_{2k} \cdots b_{nk} = (2\omega_0)^n\sqrt{n!\,(\mu_k+\nu_k+1)_n\,\!\left(\mu_k+\frac{1}{2}\right)_{\!n}\!\left(\nu_k+\frac{1}{2}\right)_{\!n}}.
\label{eq:6.17b}
\end{equation}
and $S_n(x^2;a,b,c)$ are dual Hahn polynomials. The action of the generators $K^-$ and $K^+$ on wave functions $R_n$ are defined as follows:
\begin{equation}
K^-R_{nk} = \kappa_{nk} R_{n-1,k}, \quad K^+R_n = \kappa_{n+1,k}R_{n+1,k}, \quad \kappa_{nk} = \sqrt{n(n+\mu_k+\nu_k+1)}.
\label{eq:6.18}
\end{equation}

\section{Discussion of Numerical results}
\label{v}

The numerical results are obtained by fixing all model parameters and the calculations were evaluated numerically using the Wolfram Mathematica software package, which was also employed to generate all graphical representations presented below. Here  we will use $\hbar=c=m_{0}=\Lambda=1$, and $\omega_{0}=0.1$.

\begin{figure}[htbp]
    \centering
\includegraphics[scale=0.5]{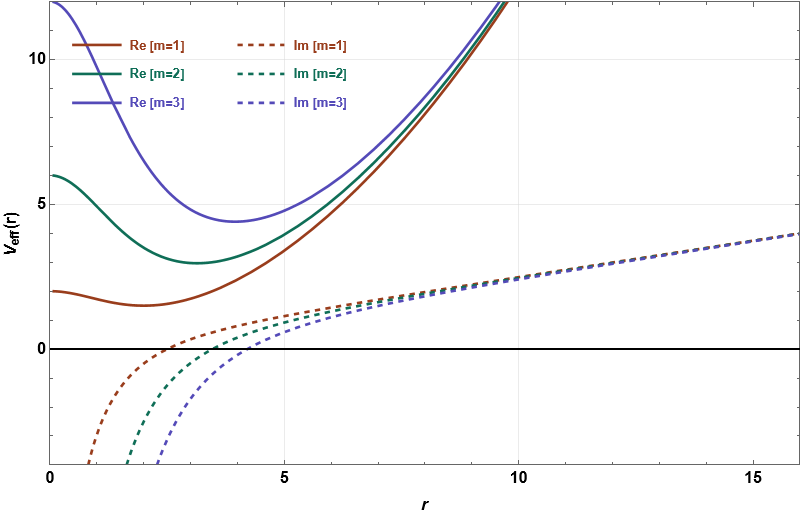}
    \hspace{10em} 
    \caption{Real and imaginary parts of the relativistic ring-shaped Quesne oscillator effective potential $V_{\mathrm{eff}}(r)$ as functions of the radial coordinate $r$ for different magnetic quantum numbers $m=1,2$ and $3$. Here, angular quantum number  $k=1$ and radial number $n=0$. Solid curves denote the real part $\mathrm{Re}\,V_{\mathrm{eff}}(r)$, while dashed curves represent the imaginary part $\mathrm{Im}\,V_{\mathrm{eff}}(r)$. 
 }
      \label{111}
\end{figure}

Figure~\ref{111} displays the effective potential of the relativistic ring-shaped Quesne oscillator  for several values of the magnetic quantum number $m$. The solid curves correspond to the real part of the potential, whereas the dashed curves represent its imaginary contribution.

The real part of the effective potential possesses a pronounced minimum, indicating the existence of a confining region capable of supporting bound states. As the magnetic quantum number increases, the minimum of the potential shifts to higher energies and larger radial distances. This behavior reflects the increasing contribution of the centrifugal-like term associated with angular motion.

The imaginary part of the effective potential originates from the finite-difference relativistic formulation and is most pronounced in the small-$r$ region. Its magnitude decreases relative to the real part as the radial coordinate increases. Consequently, the bound-state structure is primarily governed by the real component of the effective potential, while the imaginary contribution provides a characteristic relativistic correction.
 
\FloatBarrier

\begin{figure}[htbp]
    \centering

    \includegraphics[scale=0.3]{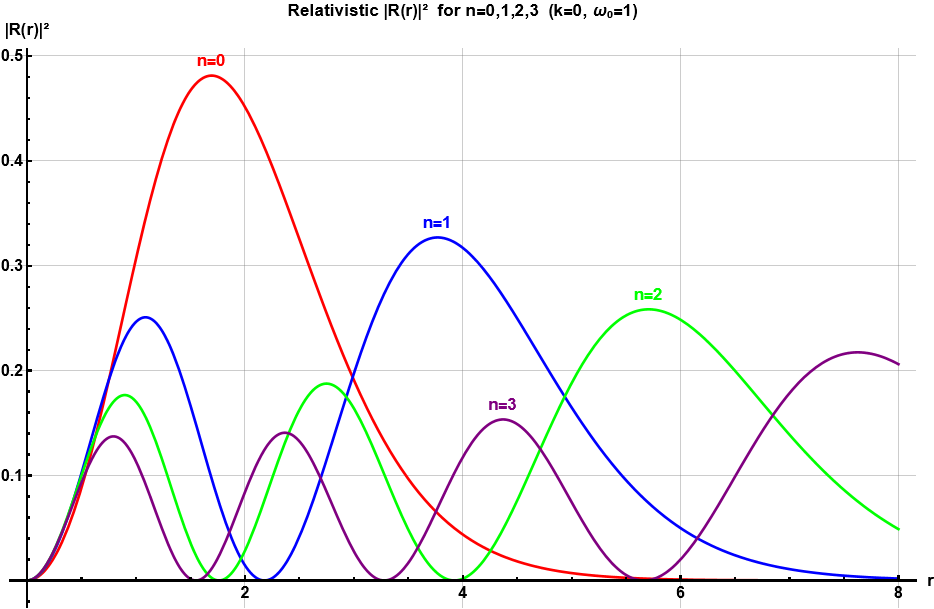}
    \includegraphics[scale=0.35]{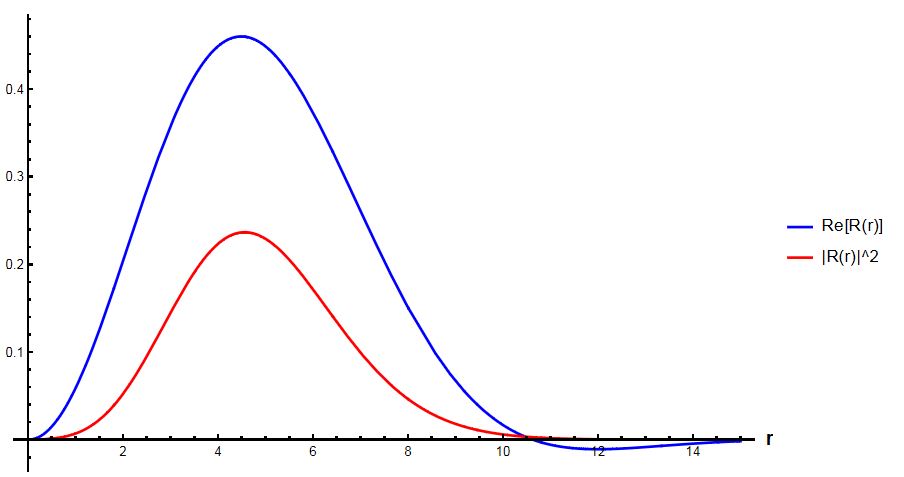}\par\textbf{$(a)\ $} \hspace{10em} \textbf{$(b)\ $}\par\vspace{2em}
   
    \hspace{1em}

 \caption{
(a) Radial probability density $|R_{nkm}(r)|^{2}$ as a function of the radial coordinate $r$ for several radial quantum numbers $n$ at fixed values of $k=0$ and $m=1$. 
(b) Comparison of the radial wave function $R_{nkm}(r)$ and the corresponding probability density $|R_{nkm}(r)|^{2}$ for the ground state ($n=0$) and $k=0$.}
    \label{fig:wavefunctions}
\end{figure}

Figure~\ref{fig:wavefunctions}(a) shows the radial probability density
$|R_{nkm}(r)|^{2}$ for several values of the radial quantum number
$n$. As the radial excitation increases, additional modes appear and
the probability distribution extends to larger values of the radial
coordinate, reflecting the larger spatial extent of excited bound
states.

For illustration, Fig.~\ref{fig:wavefunctions}(b) compares the radial wave
function $R_{nkm}(r)$ and its associated probability density
$|R_{nkm}(r)|^{2}$ for the ground state ($n=0$). The wave function
contains positive and negative regions, whereas the probability
density is nonnegative everywhere. The maximum of
$|R_{nkm}(r)|^{2}$ identifies the radial region where the particle is
most likely to be found.

\FloatBarrier

\begin{figure}[htbp]
    \centering
    \includegraphics[scale=0.3]{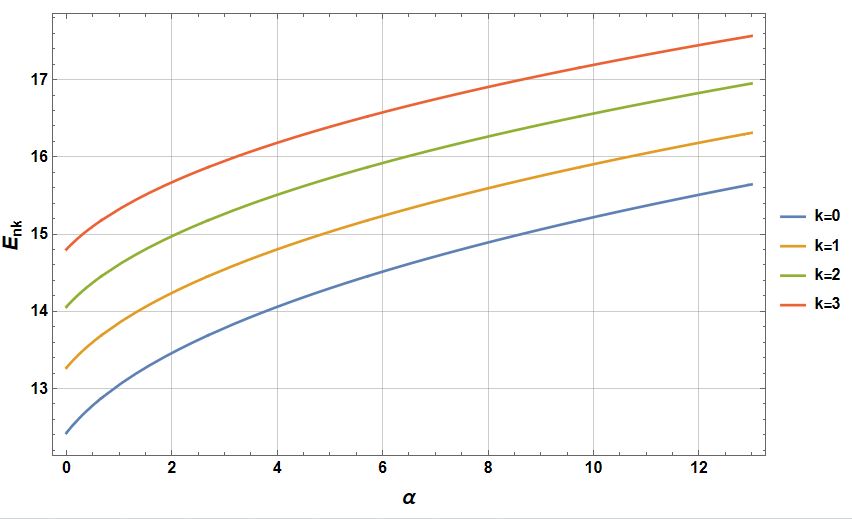}
\includegraphics[scale=0.3]{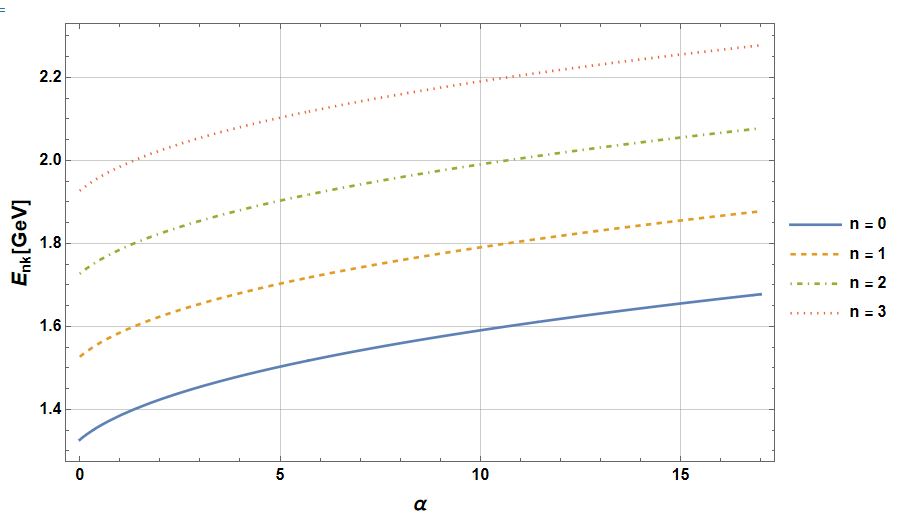}
    \par\textbf{$(a)$ } \hspace{10em} \textbf{$(b)$ }\par\vspace{2em}
    \caption{ Dependence of the relativistic energy eigenvalues $E_{nk}$ on the ring-shaped coupling parameter $\alpha$.
(a) Energy spectrum as a function of $\alpha$ for different values of the angular quantum number $k$ at fixed radial quantum number $n=0$.
(b) Energy spectrum as a function of $\alpha$ for different radial quantum numbers $n$ at fixed angular quantum number $k$.
In both cases, the energy increases monotonically with increasing $\alpha$, indicating that the ring-shaped interaction strengthens the effective confinement and shifts the bound-state spectrum toward higher energies with $k=0$.}
\label{fig:energy_alpha}
\end{figure}

The variation of the relativistic energy eigenvalues with the ring-shaped coupling parameter $\alpha$ is presented in Fig.~\ref{fig:energy_alpha}. The parameter $\alpha$ controls the strength of the noncentral ring-shaped interaction and therefore directly affects the structure of the bound-state spectrum.

Figure~\ref{fig:energy_alpha}(a) illustrates the dependence of the energy levels on $\alpha$ for several values of the angular quantum number $k$. For all considered states, the energy increases monotonically as $\alpha$ increases. Furthermore, larger values of $k$ produce systematically higher energy levels, reflecting the additional contribution of the angular part of the effective potential.

Figure~\ref{fig:energy_alpha}(b) shows the corresponding behavior for different radial quantum numbers $n$. The energy spectrum remains an increasing function of $\alpha$ for all radial excitations. States with larger $n$ are shifted upward, as expected for higher excited bound states. Although the energy separation between neighboring levels remains approximately preserved, all levels experience a noticeable upward shift with increasing $\alpha$.

These results demonstrate that the ring-shaped interaction acts as an additional confining mechanism. Increasing $\alpha$ raises the bound-state energies while preserving the ordering of the quantum states, confirming the stability of the relativistic spectrum obtained within the finite-difference formalism.
\FloatBarrier

\begin{figure}[htbp]
    \centering

    \includegraphics[scale=0.4]{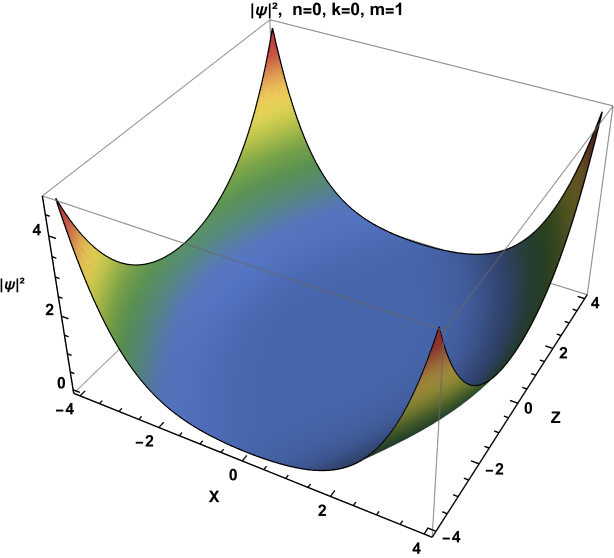} \hspace{1em}
    \includegraphics[scale=0.4]{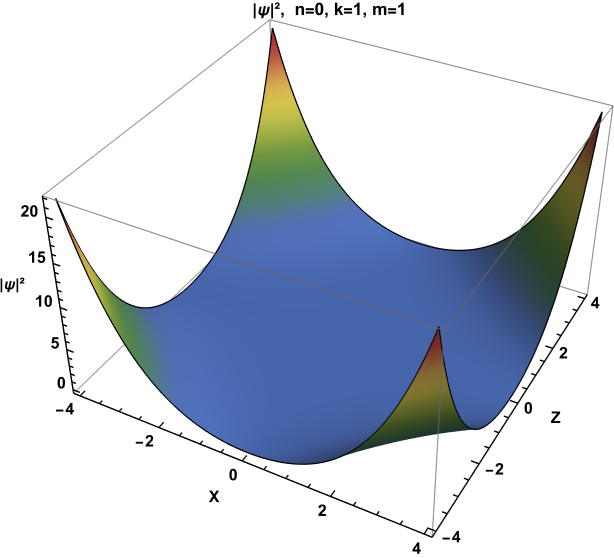}
    \par\textbf{$(a)\ k=0$} \hspace{10em} \textbf{$(b)\ k=1$}\par\vspace{2em}
    
    \includegraphics[scale=0.4]{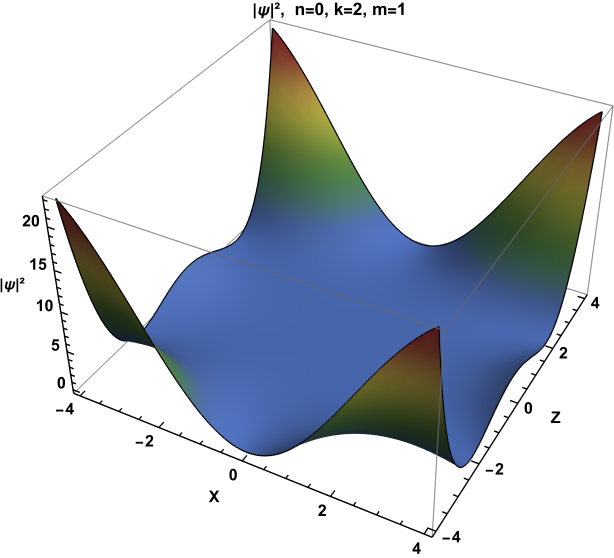}
    \includegraphics[scale=0.4]{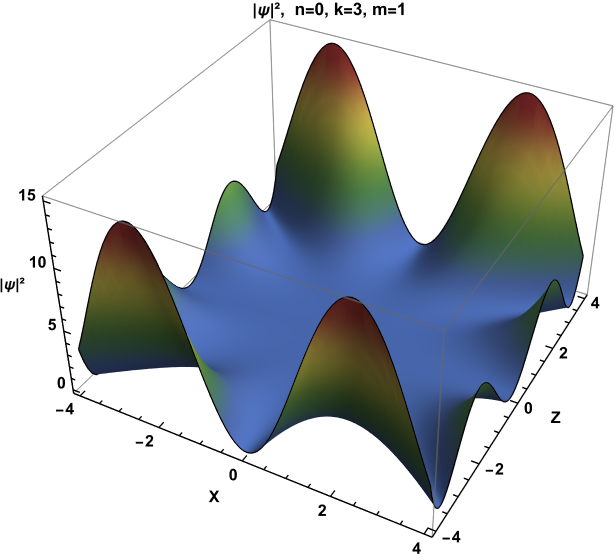}
    \par\textbf{$(c)\ k=2$} \hspace{10em} \textbf{$(d)\ k=3$}\par\vspace{2em}

    \caption{Three-dimensional spatial probability density $|\psi(x,z)|^2$ in the $X$–$Z$ plane for different values of the angular quantum number $k = 0,1,2,3$, and $n=0$.}
    \label{fig:17a}
\end{figure}

Three-Dimensional Spatial Probability Density (Fig.~\ref{fig:17a}): 
    The complete spatial behavior of the bound states within the $X-Z$ plane (In this context, we completely omit the variable $y$ from the standard 
spatial coordinate formula $r = \sqrt{x^2 + y^2 + z^2}$ (i.e., we assume 
$y = 0$). Consequently, the plot is constructed exclusively over the 
$x$ and $z$ axes. The vertical axis, on the other hand, represents the 
square of the wave function, which corresponds to the probability 
density ($|\psi|^2$)) is presented via the 3D probability density $|\psi(x, z)|^2$ in Fig.~\ref{fig:17a}. The plots reveal the explicit manifestation of quantum structures as $k$ varies from 0 to 3. For the base angular state ($k=0$), a smooth, continuous potential-like profile is observed. As the angular quantum number increases, the spatial distribution generates a well-defined sequence of localized peaks and valleys. This regular multiplication of quantum modes directly maps the structural development of higher energy bound states in the relativistic ring-shaped oscillator.

\FloatBarrier
The numerical analysis reveals that the bound-state energies increase monotonically with the ring-shaped coupling parameter $\alpha$, while the overall ordering of the quantum levels remains unchanged. Furthermore, the effective potential analysis demonstrates that the real component determines the localization and stability of the bound states, whereas the imaginary component introduces only finite relativistic corrections without altering the qualitative structure of the spectrum. These findings confirm the spectral stability of the model and highlight the role of the ring-shaped interaction in controlling the relativistic bound-state dynamics.
\section{Conclusion}
\label{con}
The application of the finite-difference relativistic quantum mechanics to a large class of physical problems requires relativistic generalizations of the exactly solvable problems of non-relativistic quantum mechanics. In this paper, we found the exact solution of the relativistic finite-difference wave equation with the ring-shaped Quesne oscillator potential. We solved the radial part and the angular part of this equation by functional method. We have shown that the radial wave function is expressed in terms of the continuous dual Hahn polynomials, and the angular wave function is expressed in terms of the Jacobi polynomials. We also have shown that the radial wave functions and the energy spectra of the model under consideration have the correct nonrelativistic limit.

As in the nonrelativistic case \cite{ref15}, the simplicity of the obtained energy spectrum suggests that a solution by group-theoretical methods should also be possible. We have shown that the radial wave equation has $SU(1,1)$ dynamical symmetry group and the angular equation has two integrals of motion $W_1$ and $W_2$. The relativistic finite-difference formulation of the ring-shaped Quesne oscillator developed in this work provides an exactly solvable framework for investigating relativistic quantum systems with non-central interactions. Owing to its analytical tractability and intrinsic relativistic structure, the model may serve as a useful platform for exploring relativistic bound states, hidden symmetries, and related classes of anisotropic quantum systems.

\end{document}